\documentclass[twocolumn,aps,prl,showpacs,superscriptaddress,amsfonts,amssymb,amsmath]{revtex4}
\usepackage{amssymb,amsmath,color}
\usepackage{graphicx}
\usepackage{graphicx}
\begin{document}
\title{Spin Transport Properties in Heisenberg
Antiferromagnetic Spin Chains:\\
Spin Current induced by Twisted Boundary Magnetic Fields}
\author{Wei Zhuo}
\affiliation{Beijing National Laboratory for Condensed matter
Physics, Institute of Physics, Chinese Academy of Sciences,
Beijing 100080, China}
\author{Xiaoqun Wang}
\affiliation{Department of Physics, Renmin University of China,
Beijing 100872, China}
\affiliation{Institute of Theoretical
Physics,Chinese Academy of Sciences,Beijing 100080,China}
\author{Yupeng Wang}
\affiliation{Beijing National Laboratory for Condensed matter
Physics, Institute of Physics, Chinese Academy of Sciences,
Beijing 100080, China}
\affiliation{International Center for
Quantum Structures, Chinese Academy of Sciences, Beijing 100080,
China}
\date{\today}

\begin{abstract}
Spin transport properties of the one-dimensional Heisenberg
antiferromagnetic spin systems for both $S=1/2$ and $S=1$ are
studied by applying twisted boundary magnetic field. The spin
current displays significantly different behavior of the spin
transport properties between $S=1/2$ and $S=1$ cases. For the
spin-half case, a London equation for the current and the
detection of an alternating electric field are proposed for the
linear response regime. The correlation functions reveal the
spiral nature of spin configuration for both ground state and the
spinon excitations. For the spin-one chain otherwise, a kink is
generated in the ground state for the size is larger than the
correlation length, leading to an exponential dependence of spin
current with respect to the chains length. The midgap state
emerges from the degenerate ground state even for small boundary
fields.
\end{abstract}


\maketitle

A significant amount of experimental and theoretical efforts has
been focused on the controlling of spin degree of freedom in
recent years.\cite{Prinz,Wolf,Zutic} For the most part the
research effort has been concentrated on the dilute magnetic
semiconductors\cite{Zutic} as well as the spin Hall effect
emerging from the spin-orbital coupling in two dimensions
\cite{MacDonald,SCZhang,Niu,ShenMaXieZhang}. However,  spin
transport properties in pure spin systems, such as the Heisenberg
system, is also of great interests because many of the novel
concepts associated with spin conduction can be tested without the
interference of charge degrees of
freedom\cite{Zotos,Fujimoto,Meisner}. For instance, ballistic
transport characterized by a finite Drude weight or spin stiffness
has been found for both the integrable systems\cite{Zotos} and
certain class of Luttinger liquids.\cite{Fujimoto} Recently, Meier
and Loss studied the magnetization transport properties in a
finite spin-half Heisenberg chain linked to two bulk
magnets.\cite{Meier} They obtained a finite spin conductivity for
a confined antiferromagnetic chain and predicted that a
magnetization current produces an electric field. Alternatively,
Sch\"utz et al\cite{Schutz} investigated a mesoscopic spin ring in
the inhomogeneous magnetic field to search for persistent spin
current for different spins. On the experimental front, mean free
paths of several hundred nanometers were found to suggest a
quasi-ballistic transport of one dimensional elementary spin
excitations in Sr$_2$CuO$_3$ and SrCuO$_2$ samples \cite{Thurber}.

In the present work, we study the spin transport properties of
one-dimensional (1D) Heisenberg antiferromagnetic (HAF) spin
models. It is well known that the integer-spin HAF chain is
distinguished by a finite gap in spectrum from the half-integer
spin one.\cite{Haldane} As expected that the spin transport is
diffusive for the integer spin case, experimental
NMR\cite{Tagigawa2} and thermal conductivity\cite{Sologubenko}
measurements have indicated finite spin diffusion and thermal
diffusion constants in AgVP$_2$S$_6$. On the other hand, Fujimoto
based on the integrability of the nonlinear $\sigma$-model
suggested that the spin transport is ballistic in the perfect 1D
spin one system.\cite{Fujimoto} Therefore, a general criteria on
ballistic spin transport are yet to be established generally for
either homogenous spin chains with different spins or mesoscopic
(quasi-)one dimensional Heisenberg systems with an inhomogeneous
magnetic field. This motivates us to devise a simplified case in
which twisted magnetic fields are applied to both ends of a finite
HAF chain with both $S=1/2$ and $S=1$. Whether the twisted
boundary magnetic fields, breaking the translational as well as
SU(2) symmetry, can properly drives the spin to flow through the
chains alternatively presents distinguishable transport nature
between the spin half and one cases.

The Hamiltonian we will consider reads:
\begin{equation}
\hat H=J\sum_{i=1}^{N-1}\hat {\bf S}_i\cdot\hat {\bf S}_{i+1}
-{\bf h}_1\cdot\hat {\bf S}_1-{\bf h}_2\cdot\hat {\bf
S}_N,\label{Hamil}
\end{equation}
where $\hat {\bf S}_i$ is the spin operator for either $S=1/2$ or
$1$ at the $i$th site, respectively. ${\bf h}_{1}$ and ${\bf
h}_{N}$ are the magnetic fields applied to the spins $\hat {\bf
S}_{1}$ and $\hat {\bf S}_{N}$. For convenience, we set $J=1$ and
$|{\bf h}_1|=|{\bf h}_N|=h$ unless specified, take ${\bf h}_1$
$=(0,0,h)$ and ${\bf h}_N$ $=$ $(h\sin\theta,0,h\cos\theta)$ where
$\theta$ is the angle between ${\bf h}_1$ and ${\bf h}_N$ and in
$[0,\pi]$. We note that such a system can be realized
experimentally by attaching two magnetic leads on two ends of the
chains.\cite{Meier}

We employ the density matrix renormalization group (DMRG)
method\cite{White,Peschel,Uli} to study the present twisted
magnetic field effects. In our computations, we use both the
infinite and the finite size algorithm. The maximum number of
sweeps is five and the number of states is kept up to 400. The
efficiency of computations is not reduced significantly by the
absence of the conserved $S_z^{total}$. The truncation errors are
about $10^{-12}$ and the relative errors maintained below one
percent as examined by increasing more kept states and sweeps. The
computations are performed up to 100 sites usually and 400 sites
for the proper extrapolations needed for thermodynamics limit.

\begin{figure}[h]
\includegraphics[width=7.5cm]{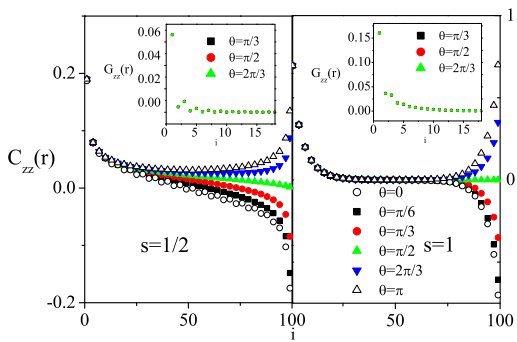}
\caption{Spin correlation function $C_{zz}(r)$ for both $S=1/2$
(left penal) and $S=1$ (right penal) for six angles $\theta=0$,
$\pi/6$, $\pi/3$, $\pi/2$, $2\pi/3 $ and $ \pi$ with $h=1$. The
insets show $G_{zz}(r)$.} \label{fig1}
\end{figure}

It is natural to first examine the twisted magnetic field
effects on the correlation function
\begin{equation}
C_{zz}(r)=(-1)^r\left<\hat S_1^z\hat S_{1+r}^z\right>
\end{equation}
for the spin $z-$component and its alternative form
\begin{equation}
G_{zz}(r)=C_{zz}(r)-(-1)^r\langle \hat S_1^z\rangle \langle
\hat S_{1+r}^z \rangle.
\end{equation}
Figure \ref{fig1} shows the correlation function $C_{zz}(r)$ with
respect to the distance $r$ for the chain length $L=100$ and
various angles at $h=1$.
We found that $C_{zz}(r)$ is
nonzero almost for all $r$ and $\theta$ for the $S=1/2$, while
$C_{zz}(r)$ is finite only when $r$ is nearby both ends of the
chain owing to gapful excitation for the $S=1$ case. This is intrinsically
associated with their algebraic and the exponent behaviors at $h=0$ for
$S=1/2$ and $1$, respectively.
As expected, $C_{zz}(r,\theta) +C_{zz}(r,\pi-\theta) =2C_{zz}(r, \frac\pi2)$.
In addition, the insets demonstrate that $G_{zz}(r)$ decays toward
zero very rapidly, surprisingly independent of the magnitude of
spins and angle $\theta$.

\begin{figure}[h]
\includegraphics[width=7.5cm]{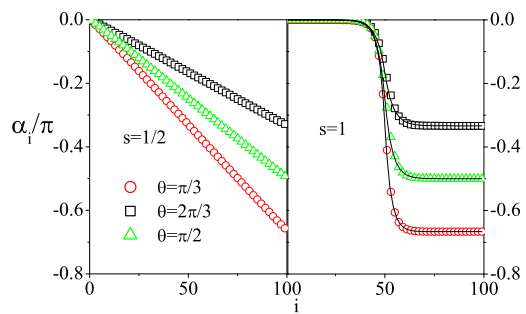}
 \caption{The spin polarization angle $\alpha_i$ as a function
$i$ with various twisted angle $\theta$ at $h=1$. In the figure,
$\alpha_i$ are explicitly displayed with odd $i$ only, while the
results for even $i$ are obtained from shifting those for odd $i$
upwards by $\pi$. The solid curves on the right panel fit Eq.
(6).}  \label{fig2}
\end{figure}

On the other hand, the twisted magnetic fields can directly alter spin
configurations in the ground state, which in principle reflects
intrinsic properties in the different spin case. We thus analyze
how the direction of
spin polarization changes with site-$i$ by introducing a classic
polarized angle defined as
\begin{equation}
\alpha_i=\tan^{-1}\left<\hat S_i^x\right>/\left<\hat S_i^z\right>
\end{equation}
which essentially measures the deviation of the spin polarization at
Figure \ref{fig2} shows $\alpha_i$ with
three different values of $\theta$ at $h=1$. For $S=1/2$, we found
that
\begin{equation}
\alpha_i=\alpha_{res}-\frac {i-1}N(\pi-\theta)+\pi \mod(i-1,2)
\end{equation}
where $\alpha_{res}$ is a residual angle which is two orders
smaller than $\alpha_i$ but depends on $i$, ${\bf h}_1$ and ${\bf
h}_N$. One can see that $\alpha_i$ depends almost linearly on $i$,
implying that ground state configuration displays a perfect
classical spiral structure in spite of strong quantum
fluctuations. The present ground state is just the superposition
of some originally lowest-lying states whose energy is the order
of the edge excitation energy and depends also weakly on the
magnetic field. Since arbitrary boundary fields do not change the
integrability of the Hamiltonian (1) for $S=1/2$, there may still
be spinon-like topological excitations\cite{cao}. However, for
$\theta\neq 0$ or $\pi$, the total spin and its $z-$component are
no longer good quantum numbers, and the spinon essentially does
not carry definite spin in contrast to the $\theta=0$ case where
the spinon possesses a spin of one half. This implies that the
spinon propagates through some kind of spiral path in response to
twisted magnetic fields. Nevertheless, since the integrability of
the system ensures the spiral spinon is dissipationless, it is
unclear what the topological charge of the spiral spinon is and
what kind of conservation law accounts for the topological charge
of the spiral spinon, which is beyond the current studies.

For the $S=1$ case, the polarization angle can be expressed with a
significantly different form
\begin{eqnarray}
\alpha_i&=&\pi \mod(i-1,2)\\
&-&\frac 1 2(\pi-\theta)(1+{\rm sign}(i-
 \frac N 2)(1-\exp({-\frac{|i-\frac N2|}{\xi_{\theta,N}}})))\nonumber
\end{eqnarray}
where ${\rm sign}(x)$ is the sign function. It is interesting to
notice that for either odd or even $i$, $\alpha_i$ exhibits a kink
at $i=N/2$ with a transient width $\xi_{\theta,N}$ which in
principle is proportional to the correlation length for
sufficiently large $N$. However, it depends on the value of
$\theta$ and independent of the magnitude of external fields as we
have checked for $h=0.1$ and $10$ cases. $\xi_{\theta,N}=2.9, 3.6$
and $4.0$ for $\theta=\pi/3, \pi/2$ and $2\pi/3$. The kink can be
interpreted as a consequence of the soliton-like nature of the magnon excitation
owing to the presence of the gap for $S=1$\cite{Haldane}.

\begin{figure}[h]
\includegraphics[width=6cm]{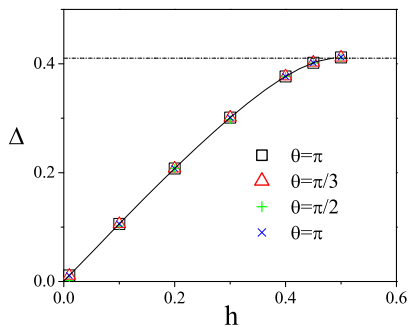}
\vspace{-0.8cm} \caption{The midgap energy versus $h$ with various
$\theta$ for the $S=1$ case.  The horizontal line indicates the
Haldane gap.\label{fig3}}
\end{figure}

For the quantum spin-1 HAF chain, there exists the $Z_2\times Z_2$ hidden
symmetry, which makes its excitations unstable against disturbances\cite{Wang97}.
It has been found that an open boundary can result in four-fold degeneracy with $S=0,1$
in the thermodynamic limit\cite{Huse}, while the magnetic doping causes
a midgap state\cite{Ditusa,Wang96}.
In the present case, the boundary field simply lifts the degeneracy and
induces subsequently a midgap. Figure \ref{fig3} shows that a midgap energy emerges linearly
in $h$ for small $h$ and enters the continuum when $h\gtrsim 0.5$. Therefore,
the boundary field can be devised as a tuning parameter to manupulate such
a macroscopic quantum state, which is detectable in neutron scattering measurements\cite{Ditusa}.
\vspace{-0.2cm}

Now we turn to the intrinsic transport properties associated with
above findings. We notice that the twisted magnetic fields essentially
impose a ``spin voltage" between two ends of spin chains\cite{Meier}. For
$S=1/2$, the spin spiral polarization homogenously extends over the whole
chain, resulting in a spin current which does not depend
on the chain length (see below). This supports
the conjuncture of Zotos and coworkers on an ideal metallic spin chain with $S=1/2$
\cite{Zotos}, resulting from the
integrability. In this sense, the correlation function $C_{zz}(r)$
in the bulk shown in Fig.\ref{fig1} properly reflects the
quasi-long range character and the edge effects near the end of
the chain. However, for the $S=1$
case, the kink prevents the propagation of the magnon
over characteristic distance $2\xi_{\theta,N}$. One can observe the spin current
only when the size is the order of $2\xi_{\theta,N}$ or smaller. This is consistent
with the conjuncture on the persist spin current on a ring proposed in
Ref.\cite{Schutz}. $C_{zz}(r)$ in Fig. \ref{fig1} indeed deplays a
short-range correlation for the spin-1 chain, while the tails
confined nearby $r\sim N$ unveils the fact that edge spins are
indeed asymptotically free due to the breaking of the hidden $Z_2\times Z_2$
symmetry.
\begin{figure}[h]
\includegraphics[width=7.5cm]{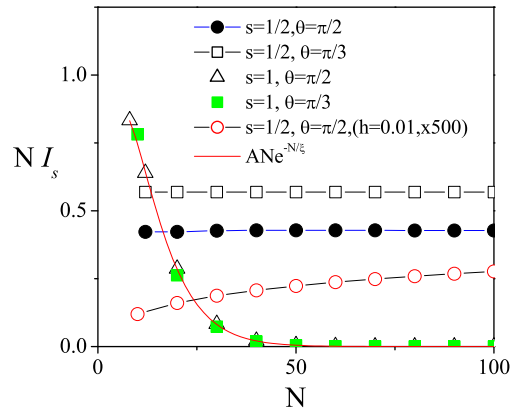}
\vspace{-0.8cm} \caption{The spin current multiplied by $N$ shows
a significantly different size-dependence for both $S=1/2$ and $1$. $h$
equals to one for various cases and is explicitly indicated
otherwise. \label{fig4}}
\end{figure}

Following Shen\cite{shen97}, we introduce a spin current:
\begin{equation}
{\bf I}_s=\left<{\bf S}_i\times{\bf S}_{i+1}\right>.
\end{equation}
which is site-independent.  As the twisted magnetic field is
applied on the $xz$ plane for both $i$ and $N$, only y-component
is non-zero. Figure \ref{fig4} shows $NI_s$ as a function of the
chain length $N$ for both $S=1/2$ and $S=1$. When $N$ increases,
it increases for small $h$ or unchange for large $h$ for $S=1/2$,
but it always decreases for $S=1$. The results for $S=1/2$
indicate again an ideal magnetic metal for the spin-half HAF
chain, while generally an insulator for $S=1$ except for small
size systems where the spin current is observable\cite{Schutz}. As
far as the spin dissipation is concerned in the $S=1$ case, we
found that $I_s =A e^{-N/\xi}$ with $A=0.39$ and $\xi=6.04$ apart
from a weak $\theta$-dependence. $\xi$ is in very good agreement
with the correlation length of the ordinary $S=1$ HAF
chain.\cite{Huse}
\begin{figure}[h]
\includegraphics[width=8.5cm]{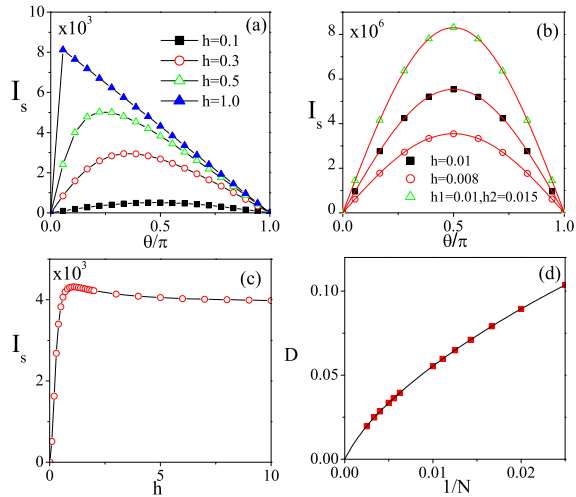}
\vspace{-0.8cm}
\caption{The spin current and the spin conductance
for $S=1/2$. (a) For various $h$ and $\theta$; (b) For the linear
 response regime with different $h_1$ and $h_N$.
 Solid curves draw $\sin\theta$,which indicate that $I_s$ is
 proportional to $h_1\times h_2$;
 (c) $I_s$ as a function of $h$  with $\theta=\pi/2$.
(d)The same size-dependence of the spin conductance with all $h$
 as the same as in (b).}
\label{fig5}
\end{figure}

Since the $S=1$ chain is an insulator, we focus on the $S=1/2$
case below. We notice that the spin current is very sensitive to
the twisted angle $\theta$ as shown in Fig.\ref{fig5}(a) and (b).
On the other hand, as seen from Fig. \ref{fig5}(c), the spin
current rapidly increases with increasing $h$ and reaches a
maximum at about $h=1$ which is the critical value of the uniform
field to saturate the uniform magnetization. When $h$ is further
increased, $I_s$ saturates. For very small ${\bf h}_1$ and ${\bf
h}_N$, the system lies in the linear response regime. In this
case, we found that the spin current is proportional to the vector
${\bf h}_1\times{\bf h}_2$ which precisely follows $\sin\theta$ as
demonstrated in Fig. \ref{fig5}(b). Therefore, we can naturally
devise a ``spin voltage" in a form of
\begin{equation}
{\bf V}_s={\bf h}_1\times{\bf h}_2,
\end{equation}
a ``spin-vector potential" for Eq. (1). Then we obtain the spin
current as follows:
\begin{equation}
{\bf I}_s=D\cdot{\bf V}_s.\label{eqD}
\end{equation}
which is one kind of the London equation with $D$ being the spin
conductance. Since both sides of the above equation are
time-reversal invariant so that the spin current is
dissipationless\cite{SCZhang}. The spin conductivity
characterizing the bulk properties described by Eq. (1) is given
by $\sigma=DN$. We calculated the spin conductance $D$ for various
${\bf h}_1$ and ${\bf h}_N$ with different chain length $N$ as
explicitly demonstrated in Fig. \ref{fig5}(d) and found that it is
linearly scaled to zero $1/N$. With making extrapolation for the
thermodynamic limit, we obtain $\sigma=10.0$ for the Heisenberg
antiferromagnetic spin-half chain, which is expected in connection
with the Drude weight studied by Zotos with the finite
temperature\cite{Zotos}.

Is the spin current discussed here observable? The system
discussed here is implementable experimentally as a two leads spin
systems introduced recently in Ref. \cite{Meier}. A spin current
generates an electric field. By measuring the electric voltage
difference between two points in the vicinity of the spin chain,
one can detect the spin current.\cite{Meier} We propose another
way of detecting the spin current in a spin half chain. With a
fixed ${\bf h}_1$ and a rotating ${\bf h}_2$, \emph{an alternating
spin current} is generated and \emph{an alternating electric
field} is subsequently observable nearby the spin chain.
Measurement of the ac voltage of a given point close to the chain
and the reference point (ground) reveals the spin current.

In conclusion, we study the spin transport properties of the
Heisenberg spin chains for $S=1/2$ as well as $S=1$ via applying
twisted boundary magnetic fields. Although the boundary conditions
generally do not affect bulk of a sufficiently large system, the
twisted boundary fields indeed change entirely spin orientation
for the chains in the thermodynamic limit, allowing to detect the
spin transport properties. The significantly different transport
properties are found for $S=1/2$ and $S=1$ chains. The former is
spin-metallic and has spiral spinon excitation, while the later is
spin-insulator which involves a static kink and unveils a midgap
state lift from the degenerate ground state by the external field.
A London-type equation for spin current with the spin voltage and
the detection of an alternating electric field are proposed for
the spin-metallic case in the linear response regime.

Authors are grateful to S.Q. Shen and X.C. Xie for fruitful
discussions. This work was supported by the National Basic
Research Program under the Grant 2005CB32170X and the NSFC Under
Grant No. 10425417 and 90203006.


\begin{references}
\bibitem{Prinz}G.A. Prinz, Science {\bf 282}, 1660 (1998).
\bibitem{Wolf}S.A. Wolf, Science {\bf 294}, 1488 (2001).
\bibitem{Zutic}I. \"Zuti\'c, J. Fabian, and S. Das Sama, Rev.
Mod. Phys. {\bf 76}, 323 (2004). References therein.
\bibitem{MacDonald}Jairo Sinova et al., Phys. Rev. Lett. {\bf 92}, 126603 (2004).
\bibitem{SCZhang}S. Murakami, N. Nagaosa, and S.C. Zhang,
 Science {\bf 301}, 1348 (2004).
 \bibitem{Niu}D. Cunova, N.A. Sinitsyn, T. Jungwirth, A.H.
MacDonald, Q. Niu, Phys. Rev. Lett. {\bf 93}, 46602(2004).
\bibitem{ShenMaXieZhang}S.Q. Shen, M. Ma, X.C. Xie and F.C. Zhang,
Phys. Rev. Lett. {\bf 92}, 256603(2004).
\bibitem{Zotos}H. Castella, X. Zotos, and P. Prelovek, Phys. Rev. Lett. 74, 972 (1995);
 X. Zotos, ibid. {\bf 82}, 1764 (1999); J. Karadamoglou and X. Zotos, ibid. {\bf 93},177203 (2004).
\bibitem{Fujimoto}S. Fujimoto, J. Phys. Soc. Jpn. 68, 2810 (1999).
\bibitem{Meisner} F. Heidrich-Meisner, A. Honecker, and W. Brenig, Phys. Rev. B 71, 184415 (2005).
\bibitem{Meier}F. Meier and D. Loss, Phys. Rev. Lett. {\bf 90},167204 (2003).
\bibitem{Schutz}F. Sch\"utz, M. Kollar and P. Kopietz, Phys. Rev.
Lett. {\bf 91}, 017205 (2003).
\bibitem{Thurber}K.R. Thurber, A.W. Hunt, T. Imai, and F.C. Chou,
Phys. Rev. Lett. {\bf 87}, 247202(2001).
\bibitem{Haldane}F.D.M. Haldane, Phys. Lett. 93A, 464 (1983).
\bibitem{Tagigawa2}M. Takigawa,  T. Asano, Y. Ajiro, M. Mekata, and Y.J. Uemura,
Phys. Rev. Lett. 76, 2173 (1996).
\bibitem{Sologubenko}A. V. Sologubenko, et al,
 Phys. Rev. B 68, 94432 (2003).
\bibitem{White}S.R. White, Phys. Rev. Lett. 69, 2863(1992).
\bibitem{Peschel}I. Peschel, X. Wang, M. Kaulke and K. Hallberg,
{\it Density Matrix Renormalization}, LNP, {\bf 528},
(1999) Springer.
\bibitem{Uli} U. Schollw\"ock Rev. Mod. Phys. 77, 259-315 (2005).
\bibitem{cao}J.P. Cao, Hai-Qing Lin, Kang-Jie Shi, and Yupeng
Wang, Nucl. Phys. B 663, 487 (2003).
\bibitem{shen97}S.Q. Shen, Phys. Lett. A{\bf 235}, 403 (1997);
S.Q.Shen and X.C.Xie, Phys. Rev. B{\bf 67}, 144423(2003).
\bibitem{Huse}S.R. White and D.A.Huse, Phys. Rev. B 48, 3844 (1993).
\bibitem{Wang97}X. Wang, Mod. Phys. Lett. B {\bf 14}, 327 (2000).
\bibitem{Ditusa}J.F. DiTusa, S.W. Cheong, J.H. Park, G. Aeppli, C. Broholm, and C.T. Chen,
Phys. Rev. Lett. {\bf 73}, 1857 (1994).
\bibitem{Wang96}X. Wang and S. Mallwitz, Phys. Rev. B {\bf 53}, R492 (1996).
\end{references}
\end{document}